\newcommand{\tpol}{\mbox{${\cal P}_{\tau}$}}

\newcommand{\gv}{\mbox{$g_V^\tau$}}
\newcommand{\ga}{\mbox{$g_A^\tau$}}
\newcommand{\gve}{\mbox{$g_V^e$}}
\newcommand{\gae}{\mbox{$g_A^e$}}
\newcommand{\gvt}{\mbox{$g_V^\tau$}}
\newcommand{\gat}{\mbox{$g_A^\tau$}}
\newcommand{\gvl}{\mbox{$g_V^\ell$}}
\newcommand{\gal}{\mbox{$g_A^\ell$}}
\newcommand{\gvabsl}{\mbox{$\mid g_V^\ell \mid^2$}}
\newcommand{\gaabsl}{\mbox{$\mid g_A^\ell \mid^2$}}
\newcommand{\gvabs}{\mbox{$\mid g_V^\tau \mid^2$}}
\newcommand{\gaabs}{\mbox{$\mid g_A^\tau \mid^2$}}

\newcommand{\slm}{\mbox{$s_L^{\tau^-}$}}
\newcommand{\stp}{\mbox{$s_T^{\tau^+}$}}
\newcommand{\stm}{\mbox{$s_T^{\tau^-}$}}
\newcommand{\snp}{\mbox{$s_N^{\tau^+}$}}

\documentstyle[aps,epsf]{revtex}  

\begin{document}        

\baselineskip 14pt
\title{Measurements of electroweak couplings of the $\tau$ lepton at L3}
\author{T. Paul}
\address{Northeastern University, Boston, MA 02115, USA}
%
\maketitle              

\begin{abstract}        

We review the current knowledge of the couplings of the $\tau$ lepton to the
electroweak gauge bosons, the W, Z and photon, obtained from the full L3
data sample at center-of-mass energies near the Z mass. Measurements of
the the effective vector and axial-vector weak neutral couplings of the $\tau$, the
Lorentz structure of the weak charged current, and anomalous couplings of
the electroweak gauge bosons to the $\tau$ are presented. 

\
\end{abstract}          

\section{Introduction}               
The sample of $\tau$ pairs produced in Z decays at LEP has been 
extensively mined as a source of data for precision electroweak
measurements as well as searches for anomalies, including
CP violation arising from sources other than the CKM matrix.
The short lifetime of the $\tau$ 
and the rather large catalog of decay modes open up analysis
possibilities which are not available for the other leptons at LEP 
energies.  In this note, we survey the measurements of $\tau$ couplings to the 
Z, the W, and the photon carried out using about $150~pb^{-1}$ of data
collected near the Z peak by the L3 detector from 1991--1995.   

\section{$\tau$--Z couplings} \label{sec:z}
Measurements of the vector and axial-vector coupling constants of the $\tau$
provide a crucial test of the hypothesis of lepton universality, and, if 
universality holds, may be combined with other electroweak data to extract a 
precise measurement of the weak mixing angle.  In addition, new 
physics, possibly including CP violation, may appear in $Z\tau\tau$ 
couplings as unexpectedly large values of weak magnetic and electric dipole moments.    

\subsection{Partial width and forward-backward asymmetry}
The partial width and forward-backward asymmetry measurements for $Z \rightarrow \tau \tau$
are described in great detail elsewhere~\cite{ref:lineshape}, and here we
reiterate only a few points.  The partial width, $\Gamma_\tau$,
is related to the effective vector and axial-vector couplings by,
\begin{equation}
\Gamma_\tau \propto \gvabs + \gaabs
\end{equation}
The forward-backward asymmetry, $A_{FB}$ is defined as,
\begin{equation}
A_{FB} = \frac{\sigma(\cos\theta>0) - \sigma(\cos\theta<0)}{\sigma(\cos\theta>0) + \sigma(\cos\theta<0)}
\end{equation}
where $\theta$ is the angle between the $\tau^-$ and beam electron.
This asymmetry is related to the coupling constants
by $A_{FB} \simeq {\cal A}_\tau {\cal A}_e$, where the polarization parameter ${\cal A}_\ell$ is 
defined by,
\begin{equation} \label{eqn:polpar}
{\cal A}_\ell = \frac {2Re(\gvl \gal^*) }{ (\gvabsl + \gaabsl) }
\end{equation}
where $\ell$ is an electron or $\tau$.

Measurements of these quantities are combined with the $\tau$ 
longitudinal polarization measurement described below to determine 
$\gv_\tau$ and $\ga_\tau$.

\subsection{$\tau$ longitudinal polarization and combined measurement of $\gvt$ and $\gat$}
\begin{figure}[ht]      
\centerline{\epsfxsize 1.5 truein \epsfbox{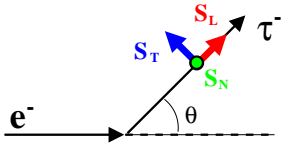}}   
\vskip -.01 cm
\caption[]{
\label{fig:spin}
\small Longitudinal, transverse, and normal spin components of the $\tau$.}
\end{figure}

The longitudinal, transverse, and normal spin components of the $\tau$, 
$S_L, S_T$ and $S_N$ respectively, are illustrated in Figure~\ref{fig:spin}.
The $\tau$ longitudinal polarization, ${\cal P}_\tau$, is defined by,
\begin{equation}
{\cal P_\tau} = \frac{\sigma(\slm=+1) - \sigma(\slm=-1)}{\sigma(\slm=+1) + \sigma(\slm=-1)}.
\end{equation}
In the Standard Model (SM), ${\cal P}_\tau$ is non-zero because of parity violation
in the weak neutral current.  It is most interesting to measure this quantity 
as a function of the $\tau$ production polar angle, in which case it is related
to the coupling constants by~\cite{ref:yellowbook},
\begin{equation} \label{eqn:ptaucostheta}
{\cal P}_\tau (\cos \theta) = -\frac{ {\cal A}_\tau (1 + \cos^2\theta) + 2 { \cal A}_e  \cos \theta }
                                   { (1 + \cos^2 \theta) + 2 { \cal A}_e  {\cal A}_\tau  \cos \theta}
\end{equation}
where the polarization parameters, ${\cal A}_\tau$ and ${\cal A}_e$, are defined 
in equation~\ref{eqn:polpar}.  Thus from this single measurement it is possible 
to determine almost independently the couplings of the Z to both the $\tau$ and the
electron.  Also note that the polarization parameters contain products of $\gv$ and $\ga$, which
means that, in contrast to the measurement of $\Gamma_\tau$ and $A_{FB}$, this measurement 
is sensitive to the relative sign of these coupling constants.

The strategy of the measurement is to 
use the energy and angular distributions of the decay products to infer the
longitudinal polarization based upon the assumption that $\tau$ decays proceed
via purely $V-A$ interactions.  The L3 analysis~\cite{ref:l3taupol} employs three complementary 
approaches.  First, a selection of exclusive decays is carried out, where we 
consider the channels $\tau \rightarrow \ell \nu_\tau \nu_\ell$ ($\ell = e,\mu$) and
$\tau \rightarrow h \nu_\tau$ ($h = \pi, \rho, a_1$).  The polarization for each 
channel is evaluated independently.  Second, an analysis of 
inclusive hadronic decays is performed which recovers some of the information lost
for cases in which exclusive identification in hadronic modes is not successful.
Third, the polarization is determined from  
the acollinearity between decay products for events in which there is at 
least one $\tau \rightarrow \pi \nu_\tau$ decay.
Results from the three methods are combined accounting for the correlations among
them.

\begin{figure}[htb]      
\centerline{\epsfxsize 2.5 truein \epsfbox{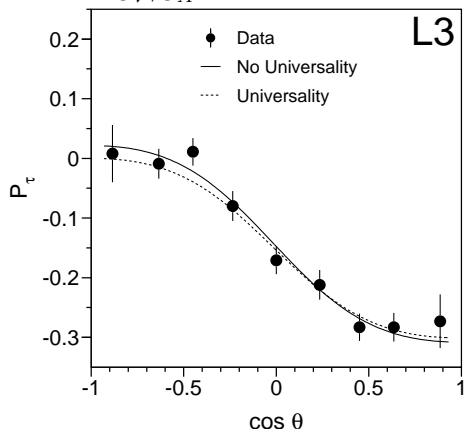}}
\vskip -.2 cm
\caption[]{
\label{fig:taupol}
\small $\tau$ polarization together with fits to equation~\ref{eqn:ptaucostheta}.}
\end{figure}

Figure~\ref{fig:taupol} shows the final measured ${\cal P}_\tau (\cos \theta)$, 
corrected for QED bremsstrahlung, $\gamma$ exchange and $\gamma$--Z interference,
together with the best fit of equation~\ref{eqn:ptaucostheta}.  The ratios of the 
effective vector and axial-vector couplings extracted from this measurement  are
$\gvt / \gat = 0.0742 \pm 0.0055$ and $\gve / \gae = 0.0845 \pm 0.0067$. This 
is consistent with lepton universality.
\vspace*{-1cm}

Assuming universality and combining the longitudinal polarization results with 
the couplings determined from $\Gamma_\tau$ and $A_{FB}$ yields the following 
preliminary results~\cite{ref:joachim}:
\begin{eqnarray*}
\gvt & = & -0.0394 \pm 0.0032 \\
\gat & = & -0.5015 \pm 0.0019 \\
\end{eqnarray*}
\subsection{Transverse-transverse and transverse-normal spin correlations}
Correlations between transverse and normal spin components of the two $\tau$'s in a 
$Z \rightarrow \tau \tau$ event
are also sensitive to $\tau$--Z couplings. In terms of the spin components depicted 
in figure~\ref{fig:spin}, transverse-transverse and transverse-normal spin correlations
are written,
\begin{eqnarray}
C_{TT} & \propto & \frac{\sigma(\stm \stp = +1) - \sigma(\stm \stp = -1)}{\sigma(\stm \stp = +1) + \sigma(\stm \stp = -1)} \\
C_{TN} & \propto & \frac{\sigma(\stm \snp = +1) - \sigma(\stm \snp = -1)}{\sigma(\stm \snp = +1) + \sigma(\stm \snp = -1)} 
\end{eqnarray}
These are related to the coupling constants by~\cite{ref:bernabeu1},
\begin{eqnarray}
C_{TT} & = & \frac{\gaabs - \gvabs}{\gaabs + \gvabs} \\
C_{TN} & = & -\frac{ 2Im(\gv \ga^*) }{\gaabs + \gvabs} + (\gamma-\mathrm{Z}~~{\mathrm Interference~~terms})
\end{eqnarray}
Although measurements of these quantities do not add much to the overall 
precision with which $\gvt$ and $\gat$ are determined, they are interesting for other
reasons.  Unlike other observables discussed so far, $C_{TT}$ is not symmetric
under the interchange of $\gvt$ and $\gat$, and is thus an on-peak observable 
which can be used to break this ambiguity.  $C_{TN}$ is of particular interest 
as it is both P and T-odd.  

\begin{figure}[ht]      
\centerline{\epsfxsize 1.5 truein \epsfbox{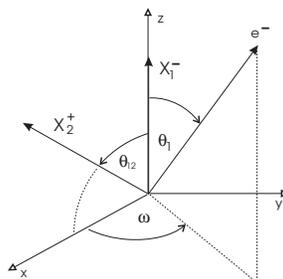}}   
\vskip -.2 cm
\caption[]{
\label{fig:omega}
\small Coordinate system used to measure spin correlations.  The $z$ axis is defined by the direction of the negatively charged $\tau$ decay product, the $x-z$ plane is defined by the directions of the two 
charged decay products.  $\omega$ is the azimuthal angle of the beam electron.}
\end{figure}

For $\tau$-pair events with exactly two charged particles in the final state, 
$C_{TT}$ and $C_{TN}$ can be measured using azimuthal asymmetries in the coordinate
system of figure~\ref{fig:omega}.
The differential cross-section in terms of $\omega$ contains 
pieces which depend on $C_{TT}$ and $C_{TN}$:
\begin{equation}
\frac{d\sigma}{d\omega} = A + B\left( {C_{TT}}\cos2\omega + {C_{TN}}\sin2\omega \right)
\end{equation}
where $A$ and $B$ are constants.

Asymmetries constructed from this distribution
yield the preliminary results~\cite{ref:spinspin}
$C_{TT} = 0.99 \pm 0.24 \pm 0.08$ and $C_{TN} = 0.15 \pm 0.24 \pm 0.07$. 
These are consistent with SM predictions~\cite{ref:bernabeu1}  of 
$C_{TT} \sim 0.99$ and $C_{TN} = -0.01$.  

\subsection{Weak dipole moments}
Weak dipole moments of the $\tau$ may be introduced through an 
effective Lagrangian which contains electric- and magnetic-type pieces,
\begin{equation}
{\cal L}^{eff}_{\mathrm{int}} = \frac{-i}{2} \frac{e {F_3^w (q^2)}}{2m_\tau} \overline{\psi}
\sigma^{\mu \nu} \gamma_5 \psi Z_{\mu\nu}
+ \frac{1}{2} \frac{e {F_2^w(q^2)}} {2 m_\tau} \overline{\psi} \sigma^{\mu\nu} \psi Z_{\mu\nu}
\end{equation}
where $Z_{\mu\nu} = \partial_\mu Z_\nu - \partial_\nu Z_\mu$. The weak 
magnetic and weak electric dipole moments are defined in terms of these
form factors as $a_\tau^w = F_2^w(q^2 = m_Z^2)$ and $d_\tau^w = e F_3^w(q^2 = m_Z^2) / 2 m_\tau$,
respectively.
In the SM, these moments are zero at tree-level but acquire 
small contributions from loops, leading to the following predictions~\cite{ref:weakloops1,ref:weakloops2}:
\begin{eqnarray*}
a_\tau^w & =    & -(2.10 + 0.61i)\times10^{-6} \\
d_\tau^w & \sim & 3\times10^{-37}e \cdot \mathrm{cm}
\end{eqnarray*}
Such tiny values are beyond the experimental reach of LEP, 
but observation of a significant non-zero value would unambiguously signal
new physics.  In particular, a non-zero value of $d_\tau^w$ would imply 
CP violation in the decay $Z \rightarrow \tau \tau$.

\begin{figure}[ht]      
\centerline{\epsfxsize 1.7 truein \epsfbox{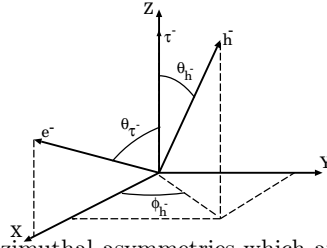}}   
\vskip -.1 cm
\caption[]{
\label{fig:weakmom}
\small Coordinate system used to define azimuthal asymmetries which are sensitive to weak
dipole moments.  Ths $z$ axis is defined by the $\tau^-$ flight direction, and the 
$x-z$ plane is defined by the $\tau^-$ flight direction and the direction of the 
beam electron.}
\end{figure}

The L3 measurement of $a_\tau^w$ and $d_\tau^w$ exploits the dependence 
of the transverse and normal polarizations of single $\tau$'s on these 
moments.  These polarization components can be determined from asymmetries
in the azimuthal angle, $\phi_h^-$, defined in figure~\ref{fig:weakmom}.
This analysis is limited to decays of the type
$e^+e^- \rightarrow \tau^+ \tau^- \rightarrow h^+\overline{\nu}_\tau h^-\nu_\tau$,
were $h^\pm$ is a charged hadron, as it is necessary to reconstruct the 
$\tau$ flight direction, a task that is not possible if there are more than two 
neutrinos in the event.
In this coordinate system, the differential cross section takes the 
form~\cite{ref:weakloops2,ref:weakloops3},
\begin{equation} \label{eqn:weakmom}
\frac{d\sigma} {d(\cos\theta_{\tau^{\pm}}) d\phi_{h^{\pm}}} = 
{\mathrm SM~part} + X \cos \phi_{h^\pm} + Y \sin \phi_{h^\pm}
\end{equation}
where $X = X(\theta_{\tau^\pm},F_2^w)$ and $Y = Y(\theta_{\tau^\pm},F_3^w)$.
Equation~\ref{eqn:weakmom} is then used to define asymmetries from which 
$a_\tau^w$ and $d_\tau^w$ can be calculated.  

%
The final results of the measurement are~\cite{ref:l3weakmom}:
\begin{eqnarray*}
Re(a_\tau^w)  & = & (0.0 \pm 1.6 \pm 2.3) \times 10^{-3} \\
Im(a_\tau^w)  & = & (-1.0 \pm 2.6 \pm 5.3) \times 10^{-3} \\
Re(d_\tau^w)  & = & (-0.44 \pm 0.88 \pm 1.33) \times 10^{-17} e \cdot \mathrm{cm}\\
\end{eqnarray*}
This is the first measurement of $a_\tau^w$.

\section{Lorentz structure of the charged current}
In addition to the studies of the weak neutral current just discussed, 
the Lorentz structure of the charged current in $\tau$ decays 
can also be studied.  Leptonic decays can be described by 
a general derivative-free four-fermion contact interaction~\cite{ref:michel1,ref:michel2},
\begin{equation}
{\cal M} = \frac{4G_F}{\sqrt{2}}
\sum_{\stackrel{\scriptstyle \gamma = S,\,V,\,T}{\lambda,\,\iota = R,\,L}}
g_{\lambda\iota}^{\gamma}
< \overline{l}_{\lambda} | \Gamma^{\gamma} | (\nu_{l})_n >
<(\overline{\nu}_{\tau})_m | \Gamma_{\gamma} | \tau_{\iota} >.
\end{equation}  
where $\iota$ and $\lambda$ label the helicity of the $\tau$ and the
final--state charged lepton respectively and $\gamma$ labels the current
as scalar, vector or tensor.  In the SM, the 
coefficient $g_{LL}^V=1$ and all the other coefficients are zero.  
The 10 possible (complex) constants $g_{\lambda \iota}^\gamma$ are conventionally expressed in terms 
of Michel parameters~\cite{ref:michel2}, of which 4 are accessible through measurements
of the leptonic decay spectra.  Specifically, the four Michel parameters
$\rho, \eta, \xi$ and $\xi\delta$ enter the decay spectrum as follows~\cite{ref:michel3}:
\begin{eqnarray} \label{eqn:michel1}
\frac{1}{\Gamma}\frac{d\Gamma}{dx_l} & = & {H_{0}^l(x_l)}-{{\cal P}_{\tau}}{H_{1}^l(x_l)}\\
  & = & {h_{0}^l(x_l)+\eta h_{\eta}^l(x_l)+\rho h_{\rho}^l(x_l)}-{{\cal P}_{\tau}}
[{\xi h_{\xi}^l(x_l)+\xi\delta h_{\xi\delta}^l(x_l)}] \nonumber
\end{eqnarray}
where $x_l$ is the lepton energy normalized to the beam energy and ${\cal P}_\tau$ 
is the $\tau$ longitudinal polarization, and the $h$'s are kinematic factors.

Semileptonic decays can be described in a similar way:
\begin{eqnarray} \label{eqn:michel2}
\frac{1}{\Gamma}\frac{d\Gamma}{dx_{h}} & = & {H_{0}^h(x_{h})}-
{{\cal P}_{\tau}}{H_{1}^h(x_{h})}\\
  & = & {h_0^h(x_{h})}-{{\cal P}_{\tau}}{\xi_{h} h_{1}^{h}(x_{h})} \nonumber
\end{eqnarray} 
where in the case of $\tau \rightarrow \rho \nu_\tau$, $x_h$ is a 
quantity constructed from the energies and angles of the $\rho \rightarrow \pi \pi^0$
decay products which maximizes sensitivity to $ {\cal P}_\tau $~\cite{ref:omega}.
The parameter $\xi_h$ is the average helicity of the $\tau$ neutrino.

From equations~\ref{eqn:michel1} and~\ref{eqn:michel2} it is apparent that
$\xi$, $\xi_h$ and $\xi \delta$ cannot be determined independently from ${\cal P}_\tau$
using just these distributions.  However, if one assumes only $V$ and $A$ interactions
in $\tau$ pair production, then the helicities of the two $\tau$'s are nearly 
$100\%$ anticorrelated.  This can be exploited to write a double decay distribution 
in which all of these Michel parameters are disentangled from ${\cal P}_\tau$:
\begin{eqnarray}
\frac{1}{\Gamma}
\frac{d^2\Gamma}{dx_Adx_B} & = &
H_0^{(A)} (x_A) H_0^{(B)} (x_B) +
H_1^{(A)} (x_A) H_1^{(B)} (x_B) \\
&  & \mbox{}- { \tpol}\   \left[
H_1^{(A)} (x_A) H_0^{(B)} (x_B) +
H_0^{(A)} (x_A) H_1^{(B)} (x_B) \right]
\end{eqnarray}

The strategy adopted for the L3 measurement is thus to carry out common 
fit to all the leptonic and semileptonic channels, including both 
joint distributions and single distributions in cases where one $\tau$ 
decay is not identified.  The Michel parameters and $\tau$ polarization are 
extracted simultaneously from this fit, with the results~\cite{ref:michel4} shown in 
Table~\ref{tab:michel}.  These results support the 
hypothesis of $V-A$ structure of the weak charged current in $\tau$ decays.

\begin{center}
\begin{table}
\caption{\label{tab:michel} Results of Michel parameter measurements and $\tau$ polarization 
together with parameters expected for purely $V-A$ interactions.}
\begin{tabular}{|c|c|c|}   \hline

Parameter      &         Measured Value              & SM expectation \\ \hline
            
$\rho$         &     0.762    $\pm$  0.035  & 0.75    \\
$\eta$         &     0.27     $\pm$  0.14   & 0       \\
$\xi$          &     0.70     $\pm$  0.16   & 1       \\
$\xi\delta$    &     0.70     $\pm$  0.11   & 0.75    \\
$\xi_h$        &    -1.032    $\pm$  0.031  & -1      \\
${\cal P}_\tau$&    -0.164    $\pm$  0.016  &         \\ \hline 
\end{tabular}
\end{table}
\end{center}

\section{Anomalous electromagnetic moments of the $\tau$}
Finally we turn to $\tau-\gamma$ couplings and in particular
what can be learned from studying radiative $\tau$ pair production
at LEP.  A $\tau$ may couple to a photon through its charge, 
magnetic dipole moment or electric dipole moment \footnote{We neglect
possible anapole moments.}.  These couplings can be parametrized by a matrix
element in which the usual $\gamma^\mu$ describing the current is replaced
by a more general Lorentz-invariant form:
\begin{equation} \label{eqn:ttg}
\Gamma^\mu = F_1(q^2)\gamma^\mu + \frac{iF_2(q^2)}{2m_\tau} \sigma ^{\mu \nu} q_\nu
- F_3(q^2) \sigma ^{\mu \nu} \gamma^5 q_\nu
\end{equation}
If $q^2 = 0$ and if the $\tau$ is real on both sides of the $\tau \tau \gamma$ vertex, 
the form factors $F_i(q^2)$ have the following interpretations: 
$F_1(0) \equiv Q_\tau$ is the electric charge; 
$F_2(0) \equiv a_\tau$ is the anomalous magnetic moment
($a_\tau \equiv (g_\tau-2)/2$); 
and 
$F_3(0) \equiv d_\tau/Q_\tau$, where $d_\tau$ is the  
electric dipole moment.
In the SM $a_\tau$ is non-zero due to loops and is calculated to be
$a_{\tau_{SM}} = 0.001\,177\,3(3)$~\cite{ref:ano1}.
Although this value turns out to be beyond the current experimental reach 
using radiative $\tau$ decays, phenomena such as
compositeness or leptoquarks~\cite{ref:comp1}
could influence values of anomalous moments.
The measurement of $d_\tau$ is also quite interesting 
as a non-zero value is forbidden by both $P$ and $T$ invariance.

In order to assess the effects of anomalous electromagnetic moments 
on radiative $\tau$ pair production, a tree level calculation of the 
squared matrix element for the process $e^+ e^- \rightarrow \tau^+ \tau^- \gamma$
has been carried out~\cite{ref:ttgtheory} which includes the contributions from 
all the form factors in equation~\ref{eqn:ttg}, Z and $\gamma$ exchange,
$\gamma$-Z interference and the interference between SM 
and anomalous amplitudes.  Large values of $F_2(0)$ or $F_3(0)$ turn out to
increase the overall rate of $\tau \tau \gamma$ events and in particular enhance the 
production of high energy, isolated photons.

\begin{figure}[ht]      
\centerline{\epsfxsize 2.0 truein \epsfbox{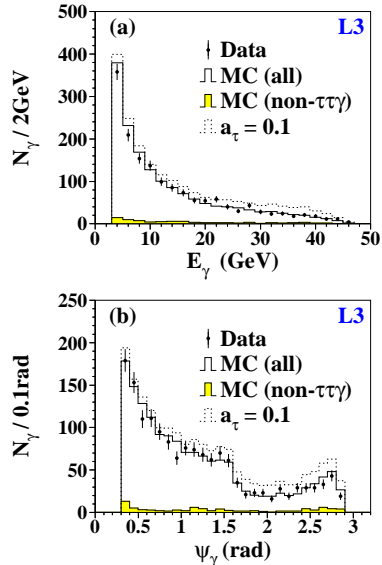}}   
\vskip -.2 cm
\caption[]{
\label{fig:ttg}
\small Number of photon candidates $(N_\gamma)$ as a function of a) photon 
energy and b) opening angle between the $\gamma$ and the nearest $\tau$.
The dashed line shows what one would expect to observe for $a_\tau = 0.1$.}
\end{figure}

Selection of $\tau \tau \gamma$ events is relatively straightforward, though
managing the background requires care~\cite{ref:l3ttg}.
Figure~\ref{fig:ttg} shows the energy and angular distributions of the selected 
events together with the Monte Carlo predictions for $F_2(0) = 0$ and 
$F_2(0) = 0.1$.
To extract 
$F_2(0)$ and $F_3(0)$ from the data, a two-dimensional fit is performed to the 
photon energy and the opening angle between the photon and the nearest $\tau$.
In order to exploit the sensitivity of the overall rate to the anomalous moments,
the normalization is fixed to the integrated luminosity.  The results are 
consistent with SM expectations, and the following 95\% C.L. limits
are set~\cite{ref:l3ttg}:
\begin{eqnarray*}
-0.052 < & F_2(0) & < 0.058 \\
(-3.1  < & F_3(0) & < 3.1) \times 10^{-16} \mathrm{e} \cdot \mathrm{cm} 
\end{eqnarray*}

\section{Conclusions}
The plethora of $\tau$ physics analysis carried out using the LEP data sample 
includes important precision measurements of weak neutral and charged 
current couplings and searches for new phenomena.  The L3 collaboration
has finalized or nearly finalized most of its analyses of $\tau$ couplings.
The results show no significant deviation from the hypothesis of lepton universality or the 
$V-A$ structure of the charged current.  The spin analysis of the $\tau$
has been recently extended beyond measurement of the longitudinal 
polarization to include transverse and normal spin correlations.
Searches for anomalous couplings did not turn up anything unexpected, but 
have provided measurements of quantities for which there was little or no
previous information, including the first measurement of the $\tau$ weak magnetic 
dipole moment and the most stringent limits on the anomalous 
electromagnetic moments.



\end{document}